\documentclass[a4paper,epj,nopacs]{svjour}
\usepackage[reqno]{amsmath}
\usepackage{amssymb}

\usepackage{graphicx}
\DeclareGraphicsExtensions{.jpg,.jpeg,.eps,.ps}

\newcommand{\csw}{c_{\text{sw}}}

\newcommand{\pslash}{\not\!p}
\newcommand{\qslash}{\not\!q}
\newcommand{\Kslash}{\not\!K}
\newcommand{\kslash}{\not\!k}
\newcommand{\Pslash}{\not\!P}

\newcommand{\quarter}{\frac{1}{4}}

\newcommand{\Tr}{\operatorname{tr}}
\newcommand{\order}{\mathcal{O}}

\newcommand{\gm}{\gamma_\mu}

\newcommand{\smn}{\sigma_{\mu\nu}}
\newcommand{\snl}{\sigma_{\nu\lambda}}

\newcommand{\dmn}{\delta_{\mu\nu}}
\newcommand{\qh}{\frac{q}{2}}

\newlength{\colw}
\newcommand{\cgluons}{\cite{Bonnet:2001uh,Bowman:2004jm,Sternbeck:2005tk}}

\newcommand{\cqqg}{\cite{Skullerud:2002ge}}
\newcommand{\cstruct}{\cite{Skullerud:2003qu}}
\newcommand{\cqqgs}{\cite{Skullerud:2002ge,Skullerud:2003qu}}

\title{Quark--gluon vertex in general kinematics}

\author{Ay{\c s}e K{\i}z{\i}lers\"u\inst{1} \and Derek B.\ Leinweber\inst{1}
  \and Jon-Ivar Skullerud\inst{2} \and Anthony G.\ Williams\inst{1}}

\institute{
Centre for the Subatomic Structure of Matter, Adelaide University, 
Adelaide, SA 5005, Australia \and
School of Mathematics, Trinity College, Dublin 2, Ireland}

\begin{document}
\bibliographystyle{h-elsevier3}

%\begin{abstract}
\abstract{
We compute the quark-gluon vertex in quenched lattice QCD, in the
Landau gauge using an off-shell mean-field $\order(a)$-improved
fermion action.  The Dirac-vector part of the vertex is computed for
arbitrary kinematics. We find a substantial infrared enhancement of
the interaction strength regardless of kinematics.}
%\end{abstract}

\maketitle

\section{Introduction}
\label{sec:intro}

In recent years, substantial progress has been made in our
understanding of the nonperturbative correlation functions
(propagators and vertices) of the fundamental fields of QCD and their
relation to the phenomena of colour confinement and dynamical chiral
symmetry breaking.  It has emerged that at least in Landau gauge, a
detailed knowledge of the structure of the quark--gluon vertex is
essential for an understanding of the dynamics of quark confinement
and chiral symmetry breaking as encoded in the quark Dyson--Schwinger
equation (DSE), relating the quark propagator $S(p)$ to the gluon
propagator and the quark--gluon vertex $\Gamma_\mu(p,q)$, where $p$
and $q$ are quark and gluon momenta respectively:
\begin{multline}
%\begin{equation} 
%\begin{split}
S^{-1}(p) = i\pslash + m \\
 + \frac{4g^2}{3}\!\int\!\!\frac{d^4q}{(2\pi)^4}  
\gamma_\mu S(p\!-\!q)D_{\mu\nu}(q)\Gamma_\nu(p\!-\!q,q) \, .
%\end{split}
\label{eq:quark-dse} 
%\end{equation} 
\end{multline}
Here, 
%$\Lambda_\nu(p,q)=ig\Gamma_\nu(p,q)$ is the quark--gluon vertex,
%while
$D_{\mu\nu}(q)=P_{\mu\nu}(q)D(q^2)$ is the gluon propagator,
with $P_{\mu\nu}(q)$ the transverse projection operator.

The overall shape of the gluon propagator is now quite well
known, both from lattice QCD \cgluons\ and from studies of the coupled
ghost--gluon Dyson--Schwinger equations.  It is now clear that if this
is fed into the quark DSE together with a bare or QED-like
vertex, the resulting quark propagator will not
exhibit a sufficient degree of chiral symmetry breaking
 \cite{Fischer:2003rp,Holl:2004qn,Iida:2003xe,Fischer:2004ym}.

The quark--gluon vertex is related to the ghost sector through the
Slavnov--Taylor identity (STI),
\begin{equation}
\begin{split}
q^\mu\Gamma_\mu(p,q) &= G(q^2)
 \Bigl[(1-B(q,p+q))S^{-1}(p)\\
& - S^{-1}(p+q)(1-B(q,p+q))\Bigr] \, ,
\end{split}
\label{eq:sti}
\end{equation}
where $G(q^2)$ is the ghost renormalisation function and $B(q,k)$ is
the ghost--quark scattering kernel.  In particular, if the ghost
propagator is infrared enhanced, as both lattice
\cite{Bloch:2003sk,Sternbeck:2005tk}
and DSE studies \cite{Fischer:2002hn} indicate, the vertex will also
be so.  This provides for a consistent picture of confinement and
chiral symmetry breaking at the level of the Green's functions of
Landau-gauge QCD, where the same infrared enhancement that is
responsible for confinement of gluons, provides the necessary
interaction strength to give rise to dynamical chiral symmetry
breaking in the quark sector.

Confinement of quarks is still not fully understood in this picture,
however.  If we ignore the scattering kernel $B$ in
(\ref{eq:sti}), dimensional analysis shows that the factor in square
brackets is proportional to $q$ as $q\to0$, so
\begin{equation}
\Gamma_\mu(p,q) \sim G(q^2) \sim (q^2)^{-\kappa}\,,q\to0\,.
\end{equation}
Using the DSE result $D(q^2)\sim(q^2)^{-1+2\kappa}$, the effective
interaction in the infrared between a quark and an antiquark is
\begin{equation}
V(q) = \Gamma(0,q)D(q)\Gamma(0,q)\sim
(q^2)^{-2\kappa}(q^2)^{-1+2\kappa} = q^{-2}\,,
\end{equation}
while a linearly confining interaction would be given by $V\sim q^{-4}$.
This indicates that the quark--gluon vertex must
contain an infrared enhancement over and above that contained in the
ghost self-energy, connected with an enhancement of the scattering
kernel.  A recent DSE calculation \cite{Alkofer:2006gz}
suggests that this may indeed be the case, with the running coupling
$\alpha_{qg}$ defined from this vertex diverging quadratically in the
infrared.

%Another area where the quark--gluon vertex may be of interest is that
%of effective charges.  Although `the running coupling' is not a
%meaningful concept beyond perturbation theory, since there is no known
%way of nonperturbatively connecting two different `schemes',
%process-dependent effective charges may be defined non-perturbatively
%and be phenomenologically useful.  The interaction between quarks and
%gluons may be a natural starting point for many of the physically
%interesting processes.

In two previous papers \cqqgs\ the quark--gluon vertex was computed on
the lattice in two specific kinematics, the soft gluon
(``asymmetric'') point $q=0$, and the quark reflection (``symmetric'')
point $q=-2p$.  Those results indicate a
highly nontrivial infrared tensor structure, a result that has been
qualitatively supported by semiperturbative DSE-based calculations
\cite{Fischer:2004ym,Bhagwat:2004kj}.  However, these kinematics may
not dominate the DSE (\ref{eq:quark-dse}).  It is therefore necessary
to compute the vertex in arbitrary kinematics, as far as possible.
This is the focus of the present paper, although at this point we will
be restricting ourselves to the dominant, vector part of the vertex.
Preliminary results were reported in \cite{Skullerud:2004gp}.

The structure of this article is as follows: In sec.~\ref{sec:formalism}
we briefly recap the formalism used in these studies.  The main
results are contained in sec.~\ref{sec:results}, while in
sec.~\ref{sec:discuss} we discuss the implications and possible future
directions.  Some tree-level lattice formulae are contained in the
Appendix.

\section{Formalism}
\label{sec:formalism}

We denote the outgoing quark momentum $p$ and the outgoing gluon
momentum $q$.  The incoming quark momentum is $k=p+q$.
In the continuum, the quark--gluon vertex can be decomposed into four
components $L_i$ contributing to the Slavnov--Taylor identity and
eight purely transverse components $T_i$:
\begin{equation} 
\begin{split}
\Gamma_\mu(p,q) & = 
 \sum_{i=1}^{4}\lambda_i(p^2,q^2,k^2)L_{i,\mu}(p,q) \\
 &\phantom{=} + \sum_{i=1}^{8}\tau_i(p^2,q^2,k^2)T_{i,\mu}(p,q) \, .
\end{split}
\label{eq:decomp}
\end{equation}
The components $L_i$ and $T_i$ are given in \cqqg:
\small
\begin{align}
L_{1,\mu}  =& \gamma_\mu\,, \notag\\
L_{2,\mu}  =& -\Pslash P_\mu\,, \notag\\
L_{3,\mu}  =& -iP_\mu\,, \notag \\
L_{4,\mu}  =& -i\sigma_{\mu\nu}P_\nu\,, \\
T_{1,\mu}  =& -i\ell_\mu\,, \notag\\
T_{2,\mu}  =& -\Pslash\ell_\mu\,, \notag\\
T_{3,\mu}  =& \qslash q_\mu - q^2\gamma_\mu\,, \notag\\
T_{4,\mu}  =& -i\bigl[q^2\smn P_\nu + 2q_\mu\snl p_\nu k_\lambda\bigr]\,,
  \notag\\
T_{5,\mu}  =& -i\sigma_{\mu\nu}q_\nu\,, \notag\\
T_{6,\mu}  =& (qP)\gamma_\mu - \qslash P_\mu\,, \notag \\
T_{7,\mu}  =& -\frac{i}{2} (qP)\smn P_\nu
 - iP_\mu\snl p_\nu k_\lambda\,, \notag\\
T_{8,\mu}  =& -\gamma_\mu\sigma_{\nu\lambda}p_\nu k_\lambda
 - \pslash k_\mu + \kslash p_\mu\,,
\end{align}
\normalsize
where $P_\mu\equiv p_\mu+k_\mu$, $\ell_\mu\equiv
(pq)k_\mu-(kq) p_\mu$.  In Landau gauge, for $q\neq0$, we are only
able to compute the transverse projection of the vertex,
$\Gamma^P_\mu(p,q) \equiv P_{\mu\nu}(q)\Gamma_\nu(p,q)$, where
$P_{\mu\nu}(q) \equiv \delta_{\mu\nu}-q_{\mu}q_{\nu}/q^2$ is the
transverse projector.  Since the vertex will always be coupled to a
gluon propagator which contains the same projector, this is also the
only combination that appears in any applications.  The four functions
$L_{i,\mu}$ are projected onto the transverse $T_{i,\mu}$, giving
rise to modified form factors
\begin{alignat}{2}
\lambda'_1 &= \lambda_1 - q^2\tau_3 &\,, \qquad
\lambda'_2 &= \lambda_2 - \frac{q^2}{2}\tau_2 \, ,\\
\lambda'_3 &= \lambda_3 - \frac{q^2}{2}\tau_1 & \, ,\qquad
\lambda'_4 &= \lambda_4 + q^2\tau_4 \, .\notag
\end{alignat}
The lattice tensor structure is more complex, and
(\ref{eq:decomp}) is only recovered in the continuum.  The form
factors also receive large contributions from lattice artefacts at
tree level, so tree-level correction is required, as described in the
Appendix.

\section{Results}
\label{sec:results}

We have analysed 495 configurations on a $16^3\times48$ lattice at
$\beta=6.0$, using a mean-field improved SW action with a quark mass
$m\approx115$ MeV.  This is part of the UKQCD data set described in
\cite{Bowler:1999ae}; further details can also be found in
\cite{Skullerud:2002ge}.

The general lattice tensor structure, even for the Dirac-vector part
of the vertex alone, is quite complicated, as the tree-level
expression in eq.~(\ref{eq:tree-gen}) of the Appendix gives an
indication of.  This makes a determination of the full tensor
structure of the vertex intractable with this lattice action.
However, in the special case where both the quark and gluon momentum
vectors are chosen to be `perpendicular' to the vertex, i.e.\ if we
compute $\Gamma_\mu(p,q)$ with $p_\mu=q_\mu=0$, this structure
simplifies considerably.  There is no loss of generality provided
rotational symmetry is restored in the continuum.  We will here only
study the leading, vector part of the vertex, as this is expected to
have the cleanest signal, and tree-level lattice artefacts can be
corrected multiplicatively.  In continuum notation, we
compute
\begin{align}
\begin{split}
\frac{1}{4}\Tr\gm&\Gamma^P_\mu(p,q) = 
 \Bigl(1-\frac{q_\mu^2}{q^2}\Bigr)\lambda'_1 \\
& + \frac{2}{q^2}\Bigl[(pq)k_\mu-(kq)p_\mu\Bigr](p_\mu+k_\mu)\lambda'_2 \\
& - [k^2-p^2-(k_\mu^2-p_\mu^2)]\tau_6
\end{split} \\
=\,& \lambda'_1 - (k^2-p^2)\tau_6 \equiv \lambda'' \, .
\end{align}
\begin{figure}
\includegraphics[width=\colw]{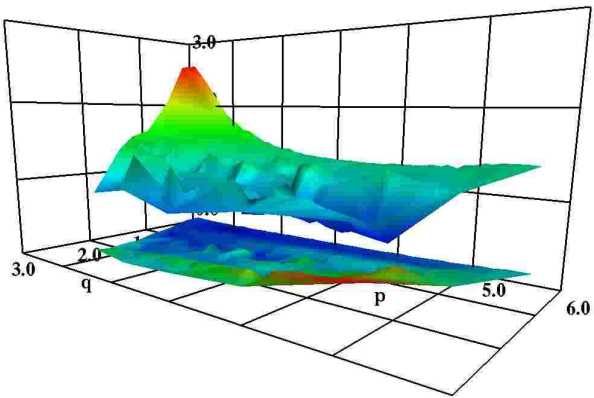}
\caption{The unrenormalised form factor $\lambda'_1$ in the
  quark-symmetric kinematics $p^2=k^2$, as a function of quark
  momentum $p$ (long axis) and gluon momentum $q$ (short axis).  The
  form factor is given by the upper surface with values on the
  vertical axis ranging from 0.63 (blue) to 2.64 (red).  The
  associated statistical uncertainties are given by the lower surface,
  with values ranging from 0.04 (blue) to 0.44 (red).}
\label{fig:qsym}
\end{figure}
Of particular interest is the quark-symmetric limit, where the two
quark momenta are equal in magnitude, $p^2=k^2$.  In this case,
$\tau_6$ is also eliminated, i.e.\ $\lambda''_1(p^2,q^2,p^2) =
\lambda'_1(p^2,q^2,p^2)$.  Note that both the soft gluon ($q=0,p=k$)
and the quark reflection ($q=-2p,p=-k$) kinematics
discussed previously in \cqqgs\ are specific instances of this more
general case.  In fig.~\ref{fig:qsym} we show $\lambda'_1$ as a
function of the two remaining independent momentum invariants.  The
data become quite noisy as $q$ in particular is increased, and also
exhibit some `spikes' and `troughs' which at present we assume to be
numerical noise and lattice artefacts.  There appears to be some
difference in the behaviour as a function of gluon momentum $q$ and
quark momentum $p$, but in view of the noise this cannot be further
quantified.

\begin{figure}
\includegraphics*[width=\colw]{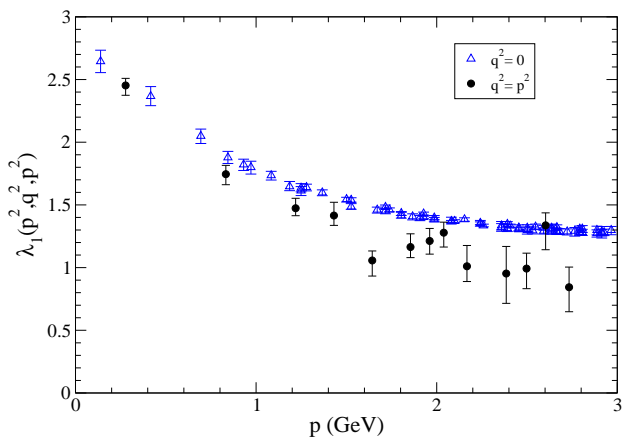}
\caption{The unrenormalised form factor $\lambda'_1$ at the totally
  symmetric kinematics $p^2=k^2=q^2$, as a function of the momentum
  $p$.  Also shown is the same form factor in the soft gluon
  kinematics $q^2=0$.}
\label{fig:symmetric}
\end{figure}

By interpolating the points in fig.~\ref{fig:qsym}, we may reach the
totally symmetric point\footnote{As we are using antiperiodic boundary
conditions in time for the fermions and periodic for the gauge fields,
it is not possible to find a totally symmetric set of lattice
momenta.} where $p^2=k^2=q^2$.  This kinematics has a history of being
used to define a momentum subtraction (MOM) scheme
\cite{Celmaster:1979km}.  We show our results in
fig.~\ref{fig:symmetric}, together with the data from the soft gluon
point presented in \cqqgs.  Again we find a strong infrared
enhancement, comparable to that of the two other kinematics (see for
example figs 7 and 8 in \cstruct).
The qualitative behaviour of $\lambda'_1$ as a function of $p^2$ is
similar in all three cases, with a tendency to drop more rapidly as
$q^2$ increases, as one would expect if the vertex is infrared
enhanced in all momentum variables.  It does not appear possible to
describe the three momentum directions with a single common variable,
i.e.\ $\lambda'_1(p^2,q^2,p^2) = \bar{\lambda}_1(t^2(p^2,q^2))$.  Note
that both sets should approach the same limit as $p^2\to0$.

\begin{figure}[tbh]
\includegraphics*[width=\colw]{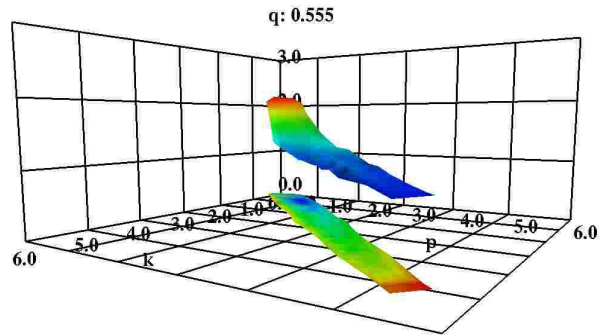}
\includegraphics*[width=\colw]{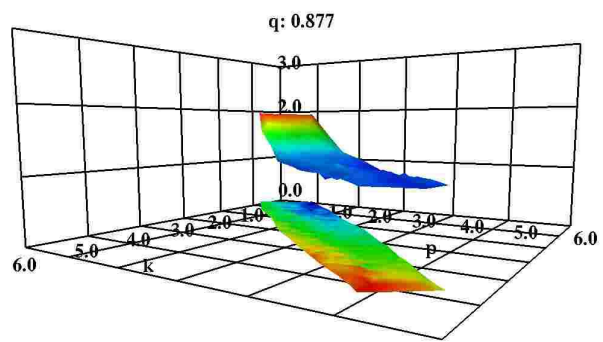}
\caption{The unrenormalised form factor $\lambda''_1$ for gluon
   momentum $q=0.555$ GeV (top) and $q=0.877$ GeV (bottom), as a
   function of quark momenta $p$ and $k$.  The values, shown by the
   upper surface, range from 1.0 (blue) to 2.2 (red), while the
   statistical uncertainties, illustrated by the lower surface, range
   from 0.03 (blue) to 0.15 (red).}
\label{fig:asym1}
\end{figure}
\begin{figure}[tbh]
\includegraphics*[width=\colw]{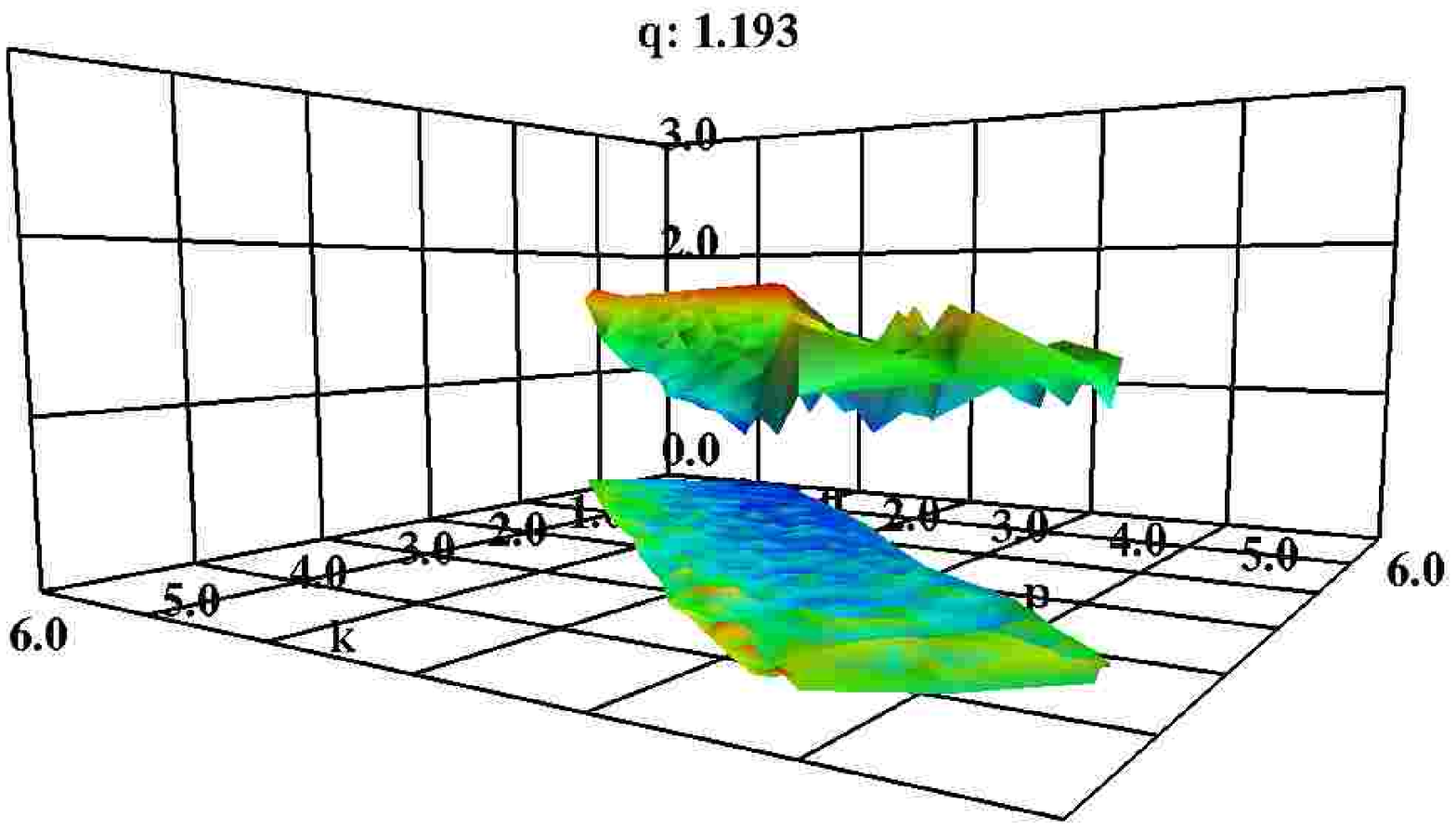}
\includegraphics*[width=\colw]{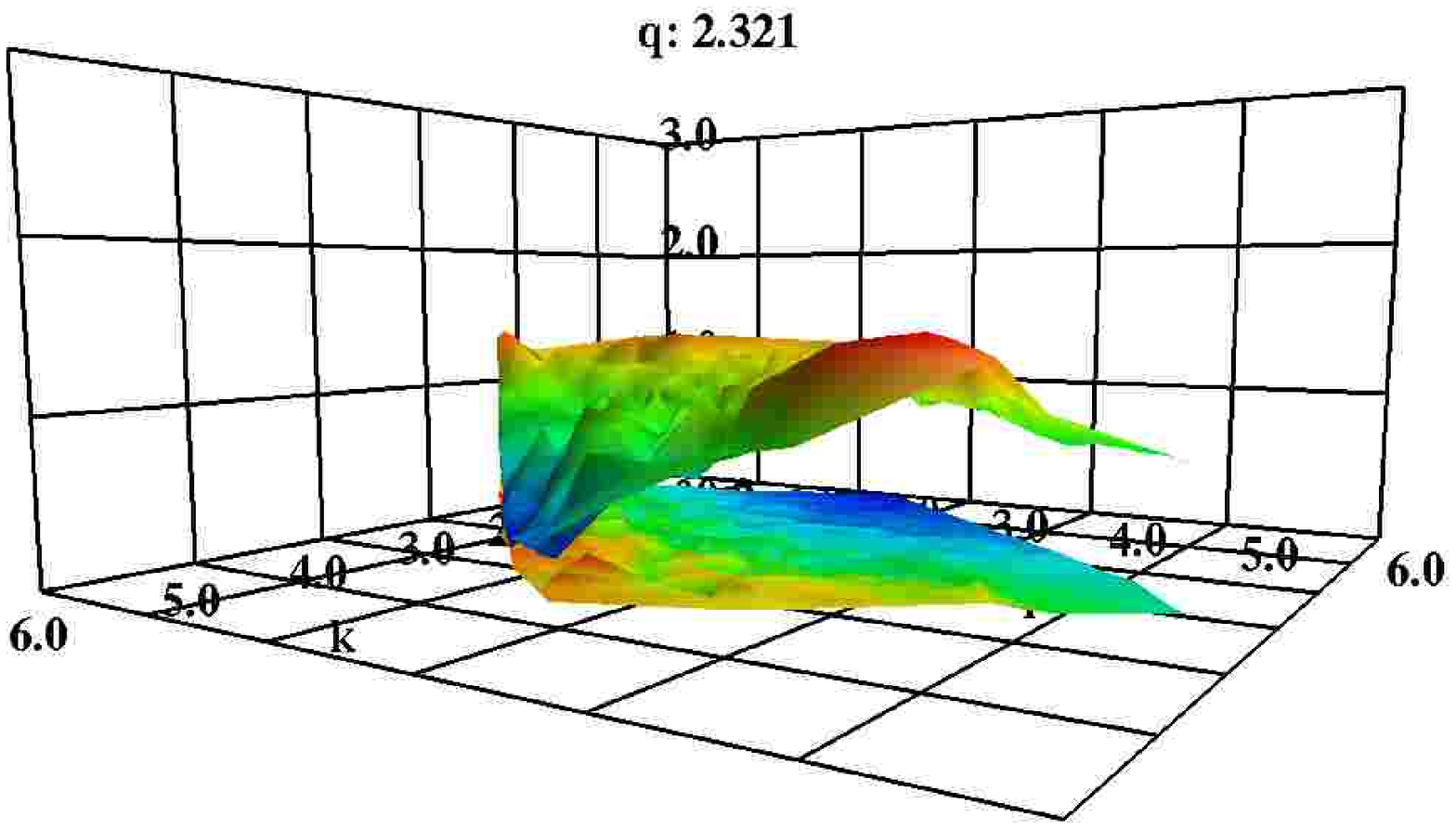}
%\vspace{-0.50cm}
\caption{As fig.~\protect\ref{fig:asym1}, but for gluon 
   momentum $q=1.193$ GeV and 2.321 GeV.  For $q=1.193$ the values
   range from 1.0 (blue) to 1.8 (red) while the uncertainties range
   from 0.03 (blue) to 0.23 (red).  For $q=1.193$ the values
   range from 0.4 (blue) to 1.5 (red) while the uncertainties range
   from 0.03 (blue) to 0.39 (red).}
\label{fig:asym3}
%\vspace{-0.60cm}
\end{figure}
Finally, figs.~\ref{fig:asym1} and \ref{fig:asym3} show $\lambda_1''$
in general kinematics, for four different fixed values of $q$, as a
function of the two quark momenta $p$ and $k$.  We expect all form
factors to be symmetric in $p^2$ and $k^2$ ($\tau_6$ on its own is
antisymmetric, but is multiplied by $p^2-k^2$), and this is also what
the figures show, within errors.  The broadening of the data surface
as $q$ grows is simply a reflection of the increase in available phase
space.

The same qualitiative features as were found in the more specific
kinematics, are reproduced here.  At low $q$, we see a clear infrared
enhancement, which disappears as $q$ grows, reflecting the fact that
at high momentum scales, only the logarithmic behaviour (which is too
weak to be seen in these data) remains.  At the same time, the level
of the surface sinks, which reflects the infrared enhancement of
$\lambda_1''$ also as a function of gluon momentum.

\section{Discussion and outlook}
\label{sec:discuss}

We have determined the leading component of the quark--gluon vertex as
a function of all three momentum invariants $p^2,k^2,q^2$.  We find an
infrared enhancement in all momentum directions, although the quality
of the current data does not make a further quantification of this
feasible.

These results have been obtained on a rather small lattice, and with a
discretisation that gives rise to quite large tree-level lattice
artefacts which must be corrected for.  We therefore expect systematic
errors to be quite large.  To obtain more reliable results, and to
extend this study to the full vertex structure at all kinematics, it
would be desirable to employ an action which is known to have smaller
and more tractable tree-level artefacts.  The Asqtad action has been
employed successfully in computing the quark propagator
\cite{Bowman:2002bm}, and unlike the SW action, only $\lambda_1$ and
possibly $\lambda_2$ are non-zero at tree level, so no tree-level
correction will be needed for the remaining form factors.  This action
is also computationally relatively cheap, making large lattice volumes
feasible.  Another possibility is using overlap fermions, which have
the advantage of retaining an exact chiral symmetry, which also
protects all the odd Dirac components of the vertex at tree level.

\begin{acknowledgement}

This work has been supported by the Australian Research Council and
the Irish Research Council for Science, Engineering and Technology.
JIS is grateful for the hospitality of the Centre for the Subatomic
Structure of Matter, where part of this work was carried out.  We
thank Patrick Bowman, Reinhard Alkofer, Christian Fischer and Craig Roberts for
stimulating discussions.
\end{acknowledgement}

\section*{Appendix}

We use the same notation as in \cstruct, defining the lattice momentum
variables
\begin{align}
K_\mu(p) & \equiv  \frac{1}{a}\sin(p_\mu a) \label{def:lat-K} \,; \\
Q_\mu(p) &
 \equiv  \frac{2}{a}\sin(p_{\mu}a/2) \label{def:lat-Q} \,; \\
C_\mu(p) & \equiv \cos(p_\mu a) \,.\label{def:lat-C}
\end{align}
The tree-level vertex can be written
\begin{multline}
\Gamma^{(0)}_{I,\mu}(p,q) = c_m
\Bigl[i\Kslash(p)A_V(p)+B_V(p)\Bigr]\times\\
\times\Bigl\{\gm C_\mu(s) - iK_\mu(s)
 - i\frac{\csw}{2}C_\mu(q/2)\sum_\nu\smn K_\nu(q)\Bigr\}\times \\
 \times\Bigl[i\Kslash(k)A_V(k)+B_V(k)\Bigr]/D_I(p)D_I(k)\,,
\label{eq:treevtx-gen}
\end{multline}
where $k=p+q; s=P/2=(p+k)/2; c_m=1+b_qma$; and $A_V, B_V$ and $D_I$ are
defined in \cstruct.  The transverse projection is given by
\begin{equation}
\Gamma^{P(0)}_{I,\mu}(p,q)
 = \sum_\nu\biggl(\dmn - \frac{Q_\mu(q)Q_\nu(q)}{Q^2(q)}\biggr)
 \Gamma^{(0)}_{I,\nu}(p,q)
\label{eq:lat-proj}
\end{equation}

Concentrating on the vector part of the vertex, we find that
\begin{equation}
\begin{split}
&\frac{D_I(p)D_I(k)}{4c_m}\Tr\Bigl(\gm\Gamma^{P(0)}_{I,\mu}(p,q)\Bigr) \\
 =&\,C_\mu(s)\biggl(1-\frac{Q_\mu^2(q)}{Q^2(q)}\biggr)\times\\
&\phantom{--}\times \Bigl[
A_V(p)A_V(k)K(p)\!\cdot\!K(k) + B_V(p)B_V(k)\Bigr]\\
&-2A_V(p)A_V(k)C_\mu(s)K_\mu(p)K_\mu(k)\\
&+A_V(p)A_V(k)\frac{Q_\mu(q)}{Q^2(q)}\times\\
&\phantom{++}\times\sum_\nu C_\nu(s)Q_\nu(q)\Bigl[
K_\mu(p)K_\nu(k) + K_\mu(k)K_\nu(p)\Bigr]\\
&+A_V(p)B_V(k)K_\mu(s)K_\mu(p) + A_V(k)B_V(p)K_\mu(s)K_\mu(k)\\
&-\frac{K(s)\cdot Q(q)}{Q^2(q)}\Bigl[
A_V(p)B_V(k)K_\mu(p) \\
&\phantom{-\frac{K(s)\cdot Q(q)}{Q^2(q)}\Bigl.--}
+ A_V(k)B_V(p)K_\mu(k)\Bigr]Q_\mu(q)\\
&-\frac{\csw}{2}C_\mu(\qh)
\biggl\{A_V(p)B_V(k)\Bigl[K(p)\!\cdot\!K(q)\!-\!K_\mu(p)K_\mu(q)\Bigr]\\
&\phantom{-\frac{\csw}{2}}
- A_V(k)B_V(p)\Bigl[K(k)\!\cdot\!K(q)\!-\!K_\mu(k)K_\mu(q)\Bigr]\biggr\}\,.
\end{split}\label{eq:tree-gen}
\end{equation}
If we now choose to impose the condition $p_\mu=q_\mu=0$,
(\ref{eq:tree-gen}) simplifies to
\begin{equation}
\begin{split}
\quarter\Tr\Bigl(\gm&\Gamma^{P(0)}_{I,\mu}(p,q)\Bigr)
= \frac{c_m}{D_I(p)D_I(k)}\times \\
\times&\biggl\{\,A_V(p)A_V(k)K(p)\!\cdot\!K(k) + B_V(p)B_V(k)\\
&-\frac{\csw}{2}
\Bigl[A_V(p)B_V(k)K(p)\!\cdot\!K(q)\\
&\phantom{-\frac{\csw}{2}\Bigl.-}
- A_V(k)B_V(p)K(k)\!\cdot\!K(q)\Bigr]\biggr\}\,.
\end{split}\label{eq:tree-spec}
\end{equation}
The tree-level correction is carried out by dividing the quantity
obtained nonperturbatively from the lattice, corresponding to the lhs
of (\ref{eq:tree-spec}), by the expression on the rhs of
(\ref{eq:tree-spec}).

\bibliography{lattice,gluon,qcd}

\end{document}